\documentclass{amsart}
\usepackage{graphicx}
\usepackage{color}
\usepackage{xcolor}
\usepackage{amsmath}
\usepackage{amssymb}
\usepackage{amsxtra} 
\usepackage{mathrsfs} 
\usepackage{yfonts} 
\usepackage{hyperref}
\usepackage{bm} 

\begin{document}

\title{Self-similar hierarchy of coherent tubular vortices in turbulence}

\author{Tomonori Tsuruhashi} 
\address{Graduate School of Mathematical Sciences, University of Tokyo, Komaba 3-8-1 Meguro, 
Tokyo 153-8914, Japan} 
\email{tomonori@ms.u-tokyo.ac.jp}

\author{Susumu Goto} 
\address{Graduate School of Engineering Science, Osaka University, 1-3 Machikaneyama, Toyonaka, Osaka 560-8531, Japan} 
\email{goto@me.es.osaka-u.ac.jp}

\author{Sunao Oka} 
\address{Graduate School of Engineering Science, Osaka University, 1-3 Machikaneyama, Toyonaka, Osaka 560-8531, Japan} 
\email{s\_oka@fm.me.es.osaka-u.ac.jp}

\author{Tsuyoshi Yoneda} 
\address{Graduate School of Economics, Hitotsubashi University, 2-1 Naka, Kunitachi, Tokyo 186-8601, Japan} 
\email{t.yoneda@r.hit-u.ac.jp} 





\begin{abstract}
Energy transfers from larger to smaller scales in turbulence. This  energy cascade is a process of the creation of smaller-scale coherent vortices by larger ones. In our recent study (Yoneda, Goto and Tsuruhashi 2021), we reformulated the energy cascade in terms of this stretching process and derived the $-5/3$ law of the energy spectrum under physically reasonable assumptions. In the present study, we provide a quantitative verification of these assumptions by using direct numerical simulations. We decompose developed turbulence in a periodic cube into scales by using the band-pass filter and identify the axes of coherent tubular vortices by the low-pressure method. Even when the turbulent kinetic energy and its dissipation rate temporally fluctuate about their temporal means, the total length of the vortices at each scale varies little with time. This result is consistent with our assumption of the temporal stationarity on the vorticity decomposition. The present numerical analysis also shows that the hierarchy of vortex axes is self-similar in a wide range of scales, i.e.~in the inertial range and a lower part of the dissipation range and that the volume fraction occupied by the tubular vortices at each scale is independent of the scale.
\end{abstract}


\maketitle

\section{Introduction}

Since the pioneering experiments by Corrsin \cite{Corrsin-1943}, it has
been well known that turbulence is not random but consists of coherent
structures. The notion of coherent structures is powerful because it can
describe, for example, the sustaining mechanism of turbulence near a
solid wall in terms of longitudinal vortices and streaks
\cite{Hamilton-1995,Waleffe-1997}. On the other hand, small-scale
turbulence in the region away from solid walls is sustained by the
energy cascade process \cite{Tennekes-1972,Frisch-1995}, where the
kinetic energy transfers from larger to smaller scales in a
scale-by-scale manner. Since the seminal direct numerical simulations
(DNS) \cite{Kerr-1985,Hussain-1986,Yamamoto-1988} in the 1980s, the
relationship between coherent structures and the energy cascade has been
studied by numerous authors; see
Refs.~\cite{Melander-1993,Lundgren-1982,Horiuti-2008,Kerr-2013,Cardesa-2017,Doan-2018} 
for example. We
also used the concept of coherent structures to clarify the concrete
energy cascade picture to demonstrate that vortices in small scales away
from solid walls are created by approximately twice as large vortices in
turbulence in a periodic cube \cite{Goto-2008,Goto-2017}, turbulent
boundary layer \cite{Motoori-2019} and turbulent channel
flow\cite{Motoori-2021}. Although such a process that the larger-scale
vortex stretches and creates smaller-scale ones has long been proposed
\cite{Tennekes-1972}, recent DNS of developed turbulence at high 
Reynolds numbers make it possible to show that developed turbulence is
indeed composed of coherent structures at various length
scales. Furthermore, such DNS can capture concrete energy-cascading
events \cite{Goto-2008,Goto-2012} and quantify the scale-locality of
energy cascade due to vortex stretching \cite{Goto-2017,Yoneda-2021}. 

In Ref.~\cite{Yoneda-2021}, using this concrete picture
\cite{Goto-2017} of energy cascade in terms of the hierarchy of coherent
vortices, we proposed a new regularity criterion for a solution
of the Navier-Stokes equation, and reformulated the energy cascade in
developed turbulence. In particular, we derived the $-5/3$ power law of
the energy spectrum from the Navier-Stokes equation without directly
using the Kolmogorov similarity hypothesis \cite{Kolmogorov-1941}.
In Ref.~\cite{Yoneda-2021}, we decomposed 
turbulence in a periodic cube into scales by using the band-pass filter
and imposed conditions on the interaction between the scales in the
stretching process. One of the most important
assumptions on the interaction between scales is that the vorticity
at each scale is expressed by coherent vortices (see (\ref{decomp}) in \S4(c) for the
concrete expression).
By using the decomposition, we represented the interaction
between large and small scale vortices
to derive the $-5/3$ power law. 
Note that we have imposed several assumptions on
the decomposition and the interactions between the scales. These
assumptions are physically reasonable and the scale-locality of the
interaction was numerically verified in
Ref.~\cite{Yoneda-2021}. However, there is no quantitative verification
of the assumptions on the decomposition itself. In this paper, we verify the
assumptions quantitatively by DNS. More concretely, we objectively
identify the axes of coherent tubular vortices by applying
the low-pressure method \cite{Miura-1997,Kida-1998, Goto-2017}
to the scale-decomposed field. From the hierarchy of
coherent vortices obtained in this way, we provide a quantitative
verification of the assumptions in
Ref.~\cite{Yoneda-2021}. 
We emphasize that this gives not only a verification of the mathematical
assumptions but also physically meaningful knowledge.
For example, we show, in the following, that
the total length of tubular vortices at each scale is independent of the
Reynolds number and forcing, and has little temporal fluctuation. The
universality of this hierarchical structure of
coherent vortices supports the assumptions in Ref.~\cite{Yoneda-2021}.
In addition, we show that the volume fraction occupied by
tubular vortices at each scale is independent of the scale in a wide
range.
This provides a new insight on the spatial intermittency of the energy
dissipation rate in turbulence, which gives a basis for the
mathematical description of the intermittency.

\section{Numerical methods}
\subsection{Direct numerical simulations}
We consider the following Navier-Stokes equation,
\begin{eqnarray} \label{ns}
      \partial_t u + u\cdot\nabla u+\nabla p = \nu \Delta u + f, \quad
      \nabla\cdot u = 0
\qquad  \mathrm{in}  \quad [0, \infty) \times \mathbb{T}^3 
\end{eqnarray}
where $u : [0,\infty)\times \mathbb T^3 \rightarrow \mathbb R^3$ is a velocity, 
$p : [0,\infty)\times \mathbb T^3 \rightarrow \mathbb R$ is a pressure, 
$f$ is an external force,
$\nu$ is the kinematic viscosity of fluid,
and $\mathbb{T}^3 = (\mathbb{R}/2\pi\mathbb{Z})^3$. We impose a suitable initial condition on $u$. We also abbreviate $u(t, x)$ to $u(x)$, because we mainly consider the case with time fixed.

In order to investigate the influence of the type of external forces, we examine two cases with different kinds of external forces. The first force $f_I$ is time-dependent but statistically homogeneous isotropic, which is expressed in the Fourier space as
\begin{equation} \label{fi}
\widehat{f_I} (t, k) = \left\{
\begin{split}
  &\frac{P}{2E_f(t)}\:\widehat{u}(t, k) &\quad &\text{if} \ \ 0 < |k| \leq k_f, \\
  &0 &\quad &\text{otherwise}.
\end{split} \right. 
\end{equation}
Here, $\widehat{f_I}$ and $\widehat{u}$ are the Fourier coefficients of $f_I$ and $u$, respectively. The parameter $P$ denotes the energy input rate and $k_f$ the maximum forcing wavenumber, which are set as $P=0.05$ and $k_f=2.5$. In (\ref{fi}), $E_f$ is the kinetic energy in the forcing range defined by
\begin{equation}
 E_f(t) = \sum_{k \in \mathbb{Z}^3,\ |k| \leq k_f} \frac{1}{2} |\widehat{u}(t,k)|^2.
\end{equation}
The second kind of forcing $f_V$ is expressed by
\begin{equation} \label{fv}
  f_V(x)=(-\sin x_1 \cos x_2, \cos x_1 \sin x_2, 0)
\end{equation}
with $x=(x_1,x_2,x_3)$. Note that $f_V$ is steady but anisotropic forcing and the forcing wavenumber of $f_V$ is $k_f=\sqrt{2}$. 

We conduct DNS by numerically solving the Navier-Stokes equation (\ref{ns}) by the standard Fourier spectral method. The nonlinear terms are evaluated by using the fast Fourier transform, where we remove the aliasing errors by the phase shift method. For time integration of (\ref{ns}), we use the fourth order Runge-Kutta-Gill scheme.

We use the external forces described in (\ref{fi}) and (\ref{fv}). For each force, we change the Reynolds number by changing $\nu$ with $f$ fixed. We set the number $N^3$ of the Fourier modes so that the dissipative scale, i.e.~the Kolmogorov length scale $\eta(t)=\nu^{3/4}\epsilon(t)^{-1/4}$ can be resolved. Here, $\epsilon(t)$ is the spatial average of the energy dissipation rate. More concretely, in our DNS with $f_I$ and $f_V$, we impose the condition $k_{\mathrm{max}}\langle\eta\rangle=1.5$ and about $1.3$, respectively, by appropriately setting $\nu$ and $N$. Here, $k_{\mathrm{max}}=\sqrt{2}N/3$ is the maximum wavenumber and $\langle\cdot\rangle$ denotes the temporal average. For the temporal integration, we choose the time increment satisfying the CFL condition.

To evaluate the development of simulated turbulence, we use the Taylor-length based Reynolds number $R_\lambda=u'\lambda/\nu$. Here, $u'$ and $\lambda$ denote the root mean square of a component of the velocity and the Taylor length, respectively. In the case of homogeneous isotropic turbulence, we can write
\begin{equation}\label{Re}
 R_\lambda(t)
 = 
 \sqrt{\frac{20}{3\nu\epsilon(t)}}\:K(t)
\end{equation}
where $K(t)$ is the kinetic energy per unit mass\footnote{In the case with $f_V$, (\ref{fv}), the flow is accompanied by inhomogeneous anisotropic mean flow and turbulence is neither statistically homogeneous nor isotropic. Therefore, (\ref{Re}) is inaccurate, but we use it to evaluate $R_\lambda$ as we can calculate it without the temporally averaged velocity field. We used (\ref{Re}) to evaluate $R_\lambda$ with the forcing $f_V$ also in the previous studies \cite{Goto-2017,Yoneda-2021}.}. By using $256^3$, $512^3$, $1024^3$ and $2048^3$ Fourier modes, we have simulated turbulence with 
$\langle R_\lambda\rangle=130, 210, 350$ and $520$ with $f_I$, and
$\langle R_\lambda\rangle=180, 290, 500$ and $670$ with $f_V$. Recall that when $R_\lambda>140$ turbulence is considered developed \cite{Dimotakis-2000} in the sense that the forcing scale significantly $2\pi/k_f$ separates from the Kolmogorov length scale $\eta$.

\subsection{Identification of tubular vortices in different scales}

It is straightforward to numerically simulate turbulence at high Reynolds numbers by the method described in the previous subsection. However, in general, there are two difficulties when we analyze coherent vortical structures in different scales in the developed turbulence. 

The first difficulty stems from the fact that we cannot extract multiple-scale features of coherent vortices by vorticity magnitude or the second invariant (i.e.~the $Q$ value) of the velocity gradient tensor. Since smallest-scale flow structures predominantly determine the velocity gradient tensor, we only observe the smallest-scale structures in the visualizations by using the vorticity magnitude or the $Q$ value. This is the reason why we need a scale decomposition to capture the multiple-scale structures in turbulence. For this purpose, when we consider turbulence in a periodic box, a filter of the Fourier modes can be used \cite{Goto-2008,Leung-2012,Goto-2012,Goto-2017} and here we also use the technique. More precisely, to extract the hierarchy of vortices, we use the band-pass filter,
\begin{align} \label{bandpass}
  \mathbb{P}(k_c)\:g(x) 
  = 
  \sum_{k \in\mathbb{Z}^3} \,
    \chi(k,k_c) \,
    \widehat{g}(k) \,
    e^{ik \cdot x}
\end{align}
where $g$ is a function, $k_c$ is the highest wavenumber in the band, $\widehat{g}$ denotes the Fourier component of $g$, and $\chi(k,k_c)$ is the characteristic function defined by
\begin{equation}\label{chi}
 \chi(k,k_c) = \left\{ 
\begin{split}
  &1 &\quad &\text{if} \ \ k_c/2\leq |k| < k_c, \\
  &0 &\quad &\text{otherwise}.
\end{split} \right. 
\end{equation}
We use the phrase a large or small scale in accordance with the value of $k_c$ because we can extract larger (or smaller) structures with smaller (or larger) $k_c$. The band-pass filter (\ref{chi}) is simple but powerful to extract multiple-scale coherent structures in spatially periodic turbulence \cite{Goto-2017}. Incidentally, for inhomogeneous turbulence, we can use real-space filters \cite{Lozano-Duran-2016,Lee-2017,Motoori-2019,Motoori-2021} for the same purpose.

The above mentioned method can decompose a simulated turbulent velocity field into different scales. However, we encounter another difficulty to objectively identify coherent vortices at each scale. Although the simplest method is to define vortices in a given scale by the regions where the scale-decomposed vorticity magnitude is larger than a certain threshold, such a method is not objective because it requires a threshold. Several methods such as $\Delta$-method \cite{Chong-1990} and $\lambda_2$-method \cite{Jeong-1995} were proposed as threshold-free methods. Here, we employ the low-pressure method\cite{Miura-1997, Kida-1998}. This method is based on the prerequisite that the pressure takes the minimum value at the center of the swirling in a cross-section of a vortex. Although the low-pressure method was originally proposed for dissipative-scale structures, it was shown in Ref.~\cite{Goto-2017} that this method was applicable at any scale. {For completeness, in the following, we briefly explain this method.

The concrete procedure of the low pressure method is as follows. First, we search candidates to construct the axes of tubular vortices. Let $p_c$ be a pressure satisfying the Poisson equation,
\begin{equation} 
  \Delta p_c
  = 
  -\nabla (u_c \cdot \nabla u_c)
\end{equation}
with $u_c(x,k_c)=\mathbb{P}(k_c)\:u(x)$. We assume that the pressure around each grid point $a$ is expressed by
\begin{equation} \label{quad}
 p_c(x)
 = 
 \sum_{|\alpha| \leq 2} \frac{D^\alpha p_c(a)}{\alpha !} (x - a)^\alpha.
\end{equation}
Here, although we have assumed the pressure around each grid point is estimated by the second order Taylor polynomial, this assumption holds in practice due to the sufficient smoothness of $p_c$. The quadratic form $(\ref{quad})$ can be written as the following normal form by changing the coordinate system induced by a suitable transformation:
\begin{equation} \label{quad2}
 p_c 
 = 
 p_{\mathrm{min}} + \sum_{j = 1}^3 \lambda^{(j)} (x'_j - b_j)^2
\end{equation}
where $x'=(x'_1, x'_2, x'_3)$ denotes a new coordinate system and $\lambda^{(1)} \geq \lambda^{(2)} \geq \lambda^{(3)}$ are the eigenvalues of the Hessian matrix of the pressure $p_c$. We denote by $c$ a foot of a perpendicular lowered from $a$ to a line, which goes through $b$ and is parallel to the eigenvector associated with $\lambda^{(3)}$. If $\lambda^{(2)} > 0$, we regard the point $c$ as being located near the axis of a tubular vortex; otherwise we discard $c$. The practical procedure to obtain $c$ is provided in Appendix A. For each grid point $a$, we obtain the sequence $\{a_i\}$ by repeating this procedure as follows: set $a_0 = a$, apply this procedure by replacing $a$ with $a_i$ and let $a_{i + 1}$ be the obtained point corresponding to $c$. We terminate this iteration at step $m$ $(1\leq m \leq 20)$ when $|a_{m}-a_{m-1}|<0.01d$. Here, $d$ denotes the grid width. In this case, we record the point $a_{m}$ as a candidate point on a vortex axis. However, we do not regard such a point as a candidate point if there exists $i$ $(i\leq m)$ such that $|a_i-a_0|>\sqrt{3}d/2$ or if the swirl condition \cite{Kida-1998} is not satisfied. Note that the linear interpolation is used to evaluate the coefficients in (\ref{quad}) at $a_i$.

In this way, we obtain the candidates. We allocate each candidate to the nearest grid point. If there exists a grid point allocated more than one candidate, we replace these candidates with a single candidate at the mass center of them. These procedures give us the final candidates to construct the axes of tubular vortices. Next we connect them. Each candidate is connected with its nearest-neighbors in the adjacent grid cells in accordance with the direction of the eigenvector associated with each $\lambda^{(3)}$. More precisely, when we denote by $e^{(3)}$ a normalized vector obtained from such an eigenvector for the candidate $q$, we search the nearest-neighbors in the right circular cone $\{x|e^{(3)}\cdot(x-q)/|x-q|>\mathrm{cos}20^{\circ}\}$ where we choose the direction of $e^{(3)}$ so that the angle with the vorticity becomes smaller. 

Note that $d$ and $\mathrm{cos} 20^{\circ}$ are artificial parameters. We set $d$ as the numerical grid width in our DNS, which is very fine compared with coherent vortices in the inertial range and sufficiently fine even in those in the dissipative range. Since the connection between the candidates $q$ can be different for different angle conditions ($\cos20^\circ$), the number of tubular vortices may depend on it. This is why we focus on the total length of tubular vortices in the following.

These axes are composed of a connected series of line segments. We define the length of each tubular vortex as the length of the polygonal chain. After identifying the axes of tubular vortices, we calculate the length of each vortex. Then, we evaluate the total length $L(k_c)$ of the vortices in the velocity field $u_c(x,k_c)=\mathbb{P}(k_c)u(x)$. In the next section, we examine the functional form of $L(k_c)$ to show the existence of the self-similar hierarchy of tubular vortices.

\section{Numerical results}

\begin{figure}
\begin{center}
\includegraphics[bb=2 464 513 720,width=0.8\textwidth]{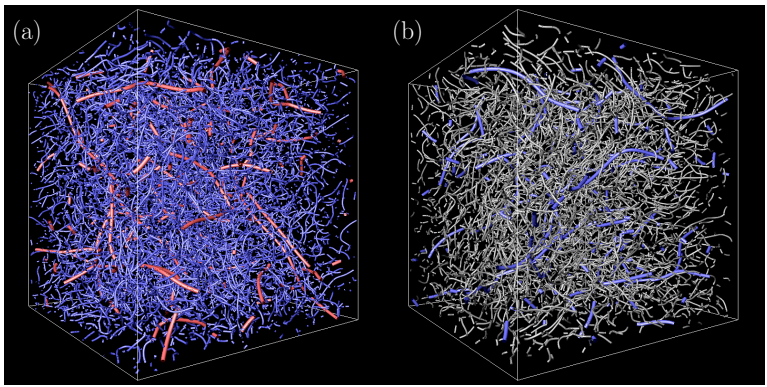} 
\end{center}
\caption{Vortex axes identified by applying the low-pressure method to the scale-decomposed velocity fields obtained by the band-pass filter (\ref{chi}) in the range $[k_c/2,k_c)$ of the turbulence ($\langle R_\lambda\rangle=350$) driven by the external force $f_I$. We show the whole domain in (a) and only a $(1/4)^3$ part of the whole domain in (b). Red curves, vortex axes for $k_c=8k_f/5$; blue, $k_c=32k_f/5$; gray, $k_c=128k_f/5$. }
\label{fig:axes}
\end{figure}


\subsection{Self-similar hierarchy of vortex axes}

First, let us observe the self-similar hierarchy of coherent vortices. We visualize in figure \ref{fig:axes} vortex axes, which we identify by the method described in the preceding section, in turbulence driven by the isotropic force $f_I$. In this figure, we show vortices at three different scales. The red curves are identified vortex axes in the largest scale ($k_c=8k_f/5$), blue curves are in the middle scale ($k_c=32k_f/5$), and gray curves are in the smallest scale ($k_c=128k_f/5$) of the three scales. Note that the red curves are at the length scale 16 times larger than the gray curves. Note also that figure \ref{fig:axes}(a) shows the whole periodic box, whereas figure \ref{fig:axes}(b) shows a $(1/4)^3$ part of the whole domain. We can see a qualitative similarity in these two panels. More specifically, the number density ratio of red to blue vortices in figure \ref{fig:axes}(a) seems similar to the ratio of blue to gray ones in figure \ref{fig:axes}(b). More quantitative arguments in the following will show that this is indeed the case. The similarity observed in these two panels implies the self-similarity of the turbulence.

\subsection{Dimension of the hierarchy}

To quantify the self-similarity demonstrated in figure \ref{fig:axes}, we show the total length $L(k_c)$ of vortex axes in the scale $\ell={k_c}^{-1}$ in figures \ref{fig:L} (a) and (b) for the turbulence driven by $f_I$ and $f_V$, respectively. In these figures, the darker symbols show the results for higher Reynolds numbers. We show the results with changing the wavenumber $k_c$ of the band-pass filter (\ref{chi}) as $k_c=1.2^i\times(8k_f/5)$ ($i=0,1,\cdots$). Note that the ratio of the lowest and highest wavenumber of the bands, $[k_c/2,k_c)$ is kept to be $2$.

\begin{figure}
\begin{center}
\includegraphics[bb=0 0 593 715,width=0.9\textwidth]{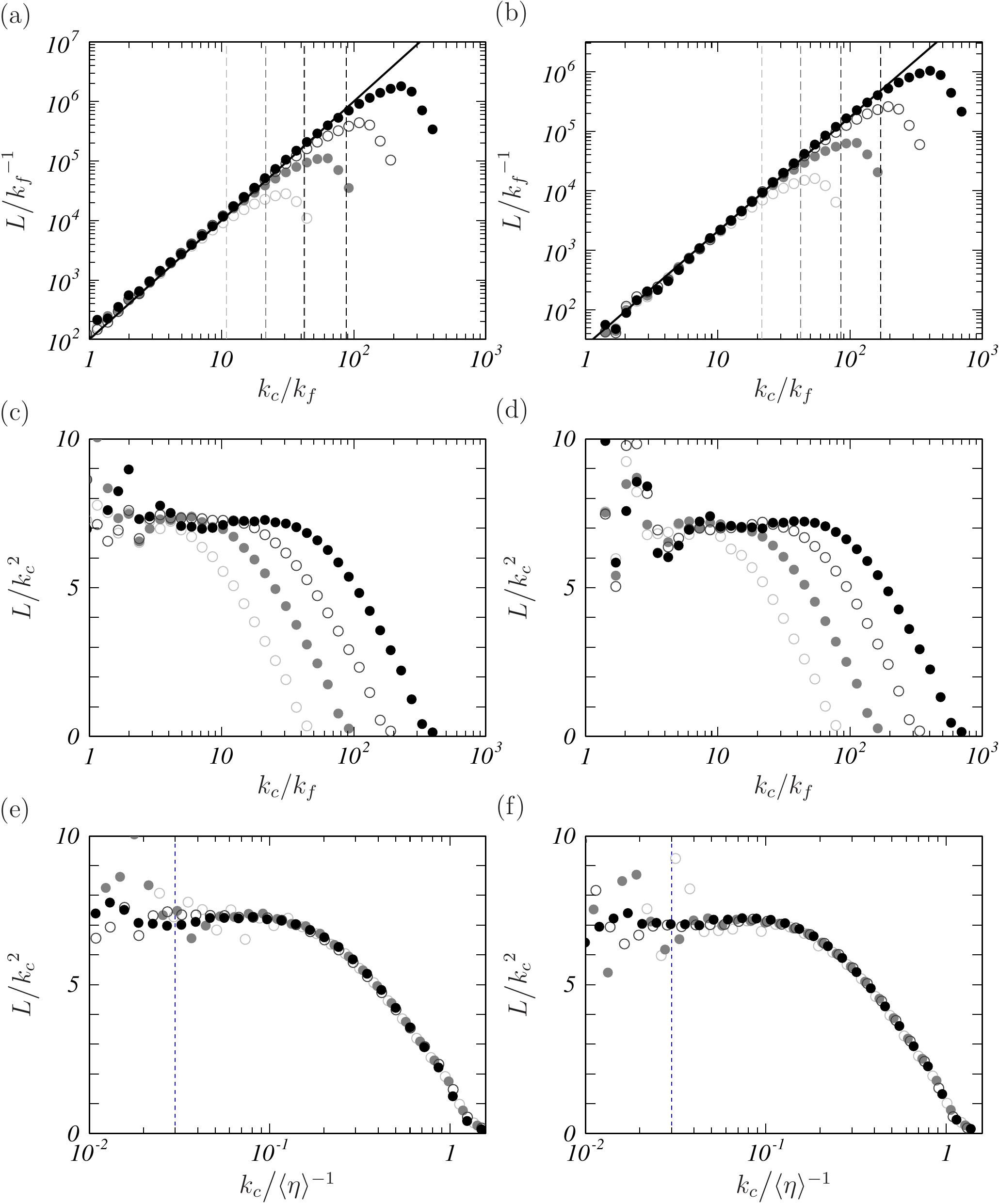} 
\end{center}
\caption{The total length $L(k_c)$ of the vortex axes in the wavenumber range $[k_c/2,k_c)$ in turbulence driven by (a, c, e) $f_I$ and (b, d, f) $f_V$. We plot, in (c--f), $L(k_c)/{k_c}^2$ as functions of $k_c$ normalized by (c, d) $k_f$ and (e, f) $\langle\eta\rangle^{-1}$. Darker symbols are for higher Reynolds numbers:
(a, c, e) $\langle R_\lambda\rangle=130$ ({\color{gray}$\circ$}),
$210$ ({\color{gray}$\bullet$}),
$350$ ($\circ$), 
$520$ ($\bullet$),
(b, d, f) $\langle R_\lambda\rangle=180$ ({\color{gray}$\circ$})
$290$ ({\color{gray}$\bullet$}),
$500$ ($\circ$), 
$670$ ($\bullet$).
Solid straight lines in (a, b) indicate the power law with the exponent being $2$. The vertical lines in (a, b) indicate $k_c=1/(3\langle\eta\rangle)$, and blue vertical lines in (e, f) indicate $k_c=0.03/\langle\eta\rangle$.}
\label{fig:L}
\end{figure}

\begin{figure}
\begin{center}
\includegraphics[bb=0 0 648 521,width=0.9\textwidth]{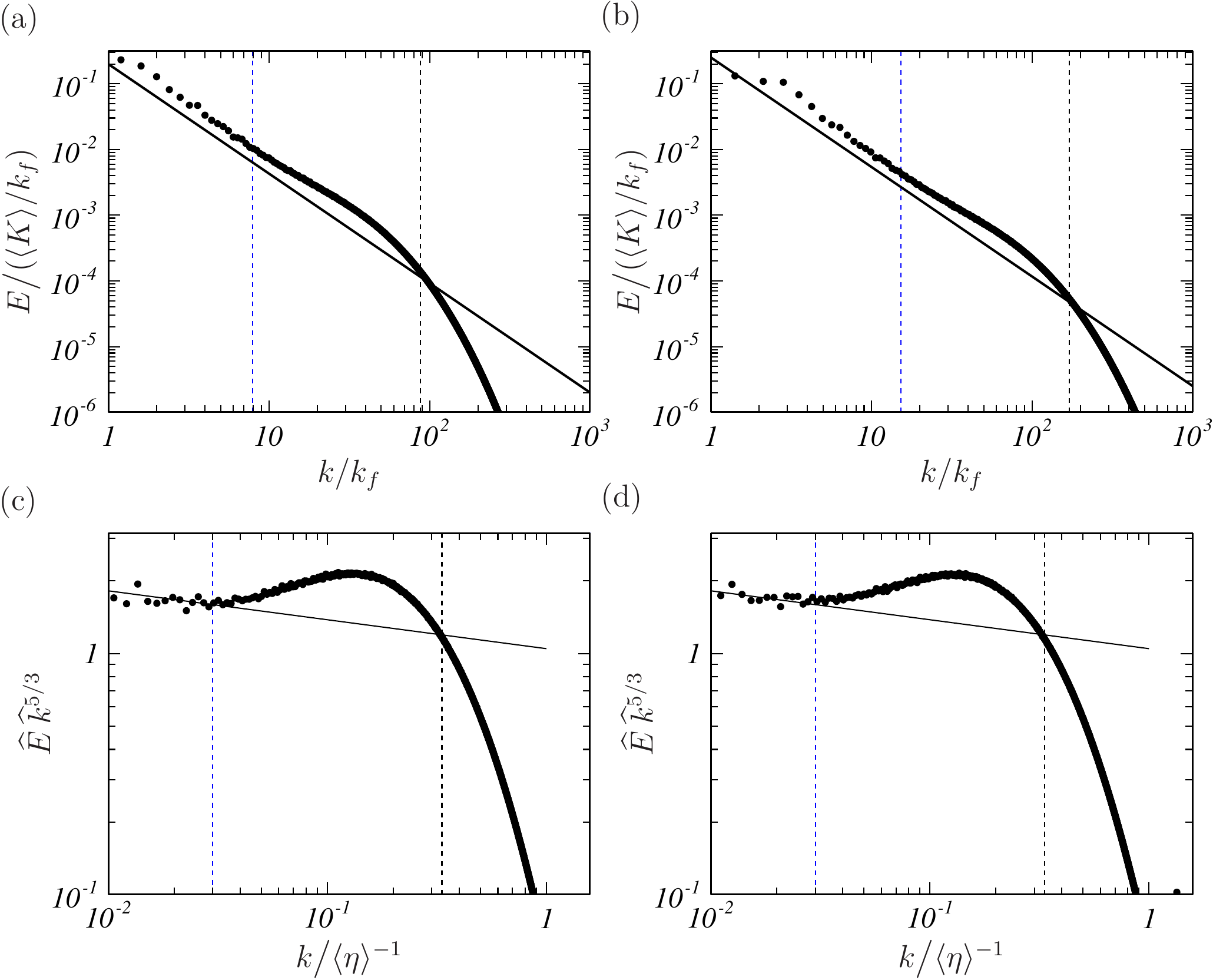} 
\end{center}
\caption{(a, b) The energy spectra $E(k)$ of turbulence driven by (a) $f_I$ and (b) $f_V$. Solid lines indicate the $-5/3$ power law, and the vertical lines are at $k=1/(3\langle\eta\rangle)$. (c, d) The compensated spectrum $\widehat{E}(k)\:\widehat{k}^{5/3}$, where $\widehat{E}=E/(\langle\epsilon\rangle^{1/4}\nu^{5/4})$ and $\widehat{k}=k/\langle\eta\rangle^{-1}$. Solid lines indicate the slope $-0.12$. In all panels, the vertical black dashed lines indicate $k=1/(3\langle\eta\rangle)$, whereas vertical blue dashed lines indicate the boundary ($k=0.03/\langle\eta\rangle$) between the T (tilted) and B (bump) ranges.}
\label{fig:E}
\end{figure}

It is remarkable in figures \ref{fig:L}(a) and (b) that $L(k_c)$ obeys a quite clear power-law function of $k_c$ in a surprisingly wide wavenumber range. We emphasize that $L(k_c)$ is estimated by a single snapshot and no temporal average is taken. Although the data deviate from a straight line in a low wavenumber range in this logarithmic plot, $L(k_c)$ seems to obey a power law for $k_c\gtrsim5k_f$. The upper cut-off wavenumber of the scaling depends on the Reynolds number. To demonstrate that it is determined by the viscous scale, we show $1/(3\langle\eta\rangle)$ for each Reynolds number by the vertical dashed lines. The data always deviate from the power law around these vertical lines, which implies that the upper cut-off wavenumber is proportional to $\langle\eta\rangle^{-1}$ irrespective of the Reynolds number and forcing. Incidentally, $L(k_c)$ rapidly decreases for $k_c\gtrsim1/\langle\eta\rangle$.

Next, we evaluate the dimension of the hierarchy of the vortices by using the statistics of $L(k_c)$. The solid straight lines in figures \ref{fig:L} (a) and (b) indicate the slope of $2$. It seems that the exponent $D_L$ of the power law is approximately $2$. For more accurate arguments we show $L(k_c)/{k_c}^2$ in figures \ref{fig:L} (c--f).  Panels (c) and (e) are results with the forcing $f_I$, whereas (d) and (f) are those with $f_V$. Looking at (c) and (d), where $k_c$ is normalized by $k_f$, $L(k_c)/{k_c}^2$ takes a constant value for $k_c\gtrsim10k_f$ in both cases of the forcing. Since a clear plateau of $L(k_c)/{k_c}^2$ is observed in this semi-logarithmic plot, we may conclude that 
\begin{equation}
 \label{eq:2.0}
 L(k_c)=C_L\:\left(\frac{k_c}{k_f}\right)^{D_L}\qquad\text{(with $C_L\approx7$ and $D_L=2$)}
 \:.
\end{equation}
Here, we assume that the radius of tubular vortices is proportional to $\ell={k_c}^{-1}$. Then, the scaling (\ref{eq:2.0}) of the total length of vortex tubes implies that the volume fraction occupied by the vortex tubes in the band $[k_c/2,k_c)$ is proportional to $L(k_c)\ell^2\propto {k_c}^0$, i.e.~the same volume fraction independent of the scale. In other words, the dimension ($D_L+1$) of the hierarchy of tubular vortices is $3$. Although this conclusion seems to be inconsistent with the classical picture that turbulent eddies distribute intermittently in space, we emphasize that we do not take into account the vortex intensity in the identification of vortices. More concretely, we use the low-pressure method, where we identify vortex axes on the basis of the local pressure distribution and streamline pattern. This is the reason why the identified vortices are always space-filling irrespective of the scale. In the energy cascading process, these vortices at each scales are intensified or weakened by the inter-scale nonlinear interactions (i.e.~vortex stretching process).

\subsection{Upper cut-off wavenumber of the hierarchy}

Here, we evaluate the upper cut-off wavenumber of the self-similarity (\ref{eq:2.0}) of the hierarchy. For this purpose, we plot $L(k_c)/{k_c}^2$ as a function of the wavenumber normalized by $\langle\eta\rangle^{-1}$ in figures \ref{fig:L} (e) and (f). We can see that $L(k_c)/{k_c}^2$ is constant for $k_c\lesssim1/(10\langle\eta\rangle)$. In summary, combining the observation in figures \ref{fig:L} (c) and (d) for the lower cut-off wavenumber, the power law (\ref{eq:2.0}) holds in the wavenumber range 
\begin{equation}
 \label{eq:range}
 10k_f\lesssim k_c\lesssim 0.1\langle\eta\rangle^{-1}.  
\end{equation}
Note that this wavenumber range (\ref{eq:range}) exists only when $R_\lambda$ is larger than about $150$. As described in \S 2, we have conducted DNS with completely different kinds of forcing; $f_I$ is statistically homogeneous and isotropic but time-dependent, whereas $f_V$ is steady but inhomogeneous and anisotropic. The shown results on the hierarchy of vortices are independent of the forcing for $k_c\gtrsim 10k_f$.

Here, we further investigate the implication of the scaling range (\ref{eq:range}) in more detail. We plot the energy spectrum $E(k)$ of the turbulence driven by $f_I$ and $f_V$ in figures \ref{fig:E} (a) and (b), respectively. For the visibility, we only plot $E(k)$ for the highest Reynolds number in each case of forcing. As widely known, the energy spectrum has a bump in wavenumbers between the inertial range, where the energy flux is constant, and the dissipation rage, where the spectrum decays exponentially. In order to compare the scaling ranges of $E(k)$ and $L(k_c)$, we plot the compensated energy spectrum $E(k)\:k^{5/3}$ in figure \ref{fig:E}(c) and (d). Ishihara {\it et al.} \cite{Ishihara-2016} showed by the DNS at much higher Reynolds numbers (up to $R_\lambda=2297$) than ours that there are distinct wavenumber ranges; namely, from the higher wavenumbers, the B (bump) range where $E$ has a bump, the T (tilted) range where $E(k)\propto k^{-5/3-\mu}$ (with the intermittency exponent $\mu\approx0.12$), and the F (flat) range where the compensated spectrum $E(k)k^{5/3}$ exhibits a plateau indicating the Kolmogorov spectrum without intermittency effects. According to Ref.~\cite{Ishihara-2016}, the boundary between the F and T ranges is about $0.005\langle\eta\rangle^{-1}$ and that between T and B is about $0.02\langle\eta\rangle^{-1}$. The energy spectrum in our DNS also shows similar behaviors (figure \ref{fig:E}), although there is no F range because of the smallness of $R_\lambda$. More specifically, the boundary between the T and B ranges is located around $0.03\langle\eta\rangle^{-1}$ (figure \ref{fig:E}(c, d)). Note that this value is slightly larger than the value $0.02\langle\eta\rangle^{-1}$ suggested in Ref.~\cite{Ishihara-2016}. Recalling the range (\ref{eq:range}) of the self-similarity of the vortex hierarchy (figure \ref{fig:L}), we conclude that the self-similarity holds in T range and a lower part of the B range.

Before closing this subsection, it is worth mentioning the relationship shown above and the energy cascading process. In Ref.~\cite{Yoneda-2021}, we have shown the self-similar energy transfer due to vortex stretching; that is, vortices at given length sales acquire the energy from about 1.7 times larger vortices, and transfer it to about $1/1.7$ times smaller vortices. We have also shown that this self-similarity is valid in the wavenumber range smaller than $0.03\langle\eta\rangle^{-1}$; namely, in the T range. In the B range, vortices transfer their energy to $1/1.7$ times smaller vortices, but the size of mother vortices changes because of the attenuation of vortices by the viscous effects; see figure 1(c) in Ref.~\cite{Yoneda-2021}.

\subsection{Steadiness of the hierarchy}

It is known that $E$ can temporally fluctuate about its temporal mean because the intensity of vortices transfers from larger scales to smaller ones \cite{Ohkitani-1992,Yasuda-2014}. In contrast, $L(k_c)$ always obeys a clear power law. To demonstrate this feature, we plot the temporal evolution of the spatially-averaged kinetic energy $K(t)$ and average dissipation rate $\epsilon(t)$ in figure \ref{fig:T-dep}(a). This is a result from the DNS with the forcing $f_V$. As was shown in Refs.~\cite{Yasuda-2014,Goto-2017}, these quantities in turbulence driven by the steady force $f_V$ evolves quasi-periodically with significant amplitude with respect to their temporal means. Within this time period, we take six instants with a fixed time interval ($\Delta t/T=4.5$) to plot $L(k_c)$ in figure \ref{fig:T-dep}(b) with six different symbols. Despite the significant fluctuations in $K(t)$ and $\epsilon(t)$, $L(k_c)$ almost perfectly steady in the high wavenumber range ($k_c\gtrsim10k_f$). Recall that the vortex axes are identified by the low-pressure method without any assumption on the vortex strength. Hence, these results imply that the number of tubular vortices are almost independent of time even if the vortex strength changes. This result supports one of the assumptions in Ref.~\cite{Yoneda-2021}, which we will discuss in more detail in the next section.

\begin{figure}
\begin{center}
\includegraphics[bb=0 0 647 264,width=0.9\textwidth]{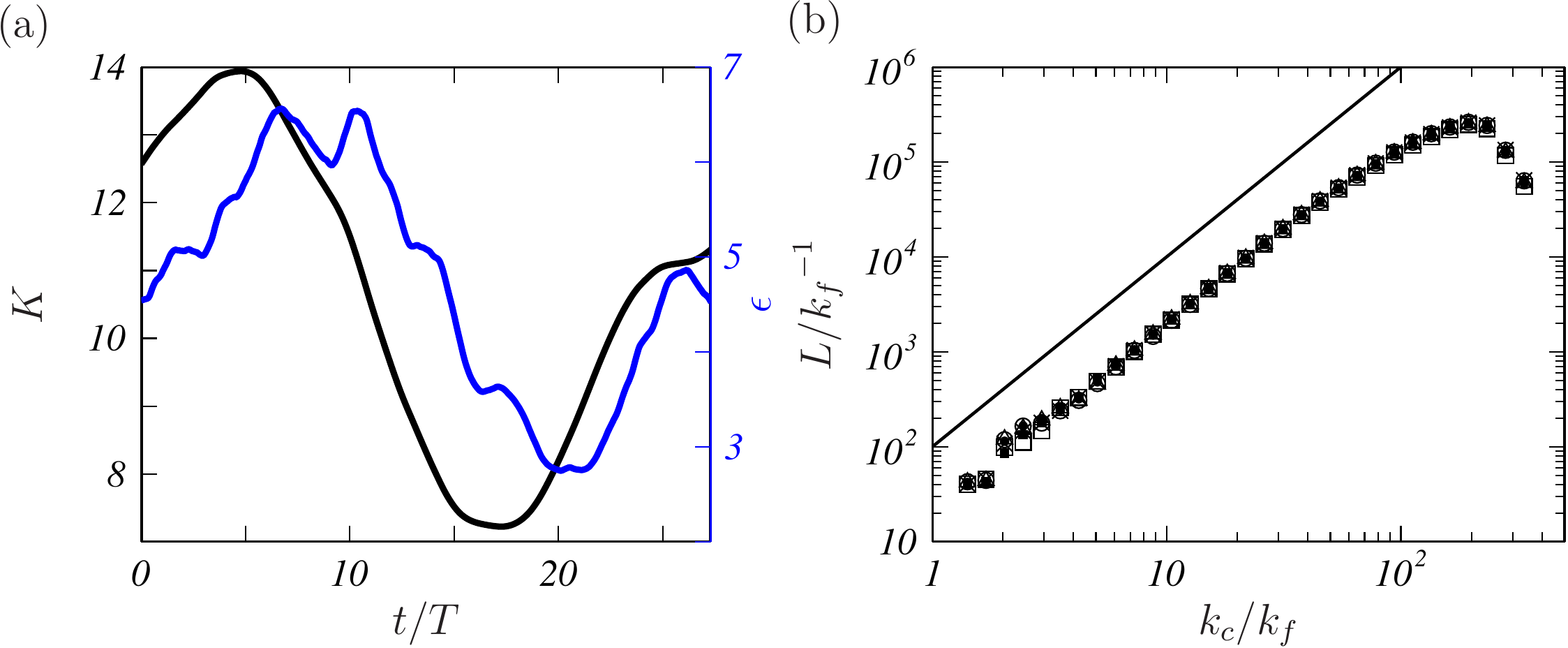} 
\end{center}
\caption{(a) Time series of the spatially averaged kinetic energy $K(t)$ and average dissipation rate $\epsilon(t)$ in the turbulence sustained by $f_V$ at the Reynolds number $\langle R_\lambda\rangle=500$. Time is normalized by the turnover time $T$ of the largest eddies. We observe a quasi-periodic behavior with a significant magnitude. (b) $L(k_c)$, normalized by ${k_f}^{-1}$ at six different instants at $t/T=4.6$, $9.1$, $14$, $18$, $23$ and $27$ in the period shown in (a). We show six data with different symbols, all of which are almost perfectly coincide for $k_c\gtrsim10k_f$. Solid line indicates the power law with the exponent $2$.}
\label{fig:T-dep}
\end{figure}

\section{Discussions}

\subsection{Justification of our energy cascade model}

On the basis of the numerical results shown in the preceding section, we discuss the assumption in Ref.~\cite{Yoneda-2021}. 
By using the notation in this paper, the vorticity decomposition, which is one of the most important assumptions in Ref.~\cite{Yoneda-2021}, can be written as follows: 
\begin{align} \label{decomp}
\mathbb{P}(k_c) \ \omega(t, x)
=
A(k_c) \sum_{j=1}^{N(k_c)} W_{j, k_c}(x)
\end{align}
where $A(k_c)$ is the amplitude of this vorticity,
$W_{j, k_c}$ is the vorticity of each tubular vortex
and $N(k_c)$ is the number of tubular vortices at this scale. 
There are three points to note in this assumption. 
First, this decomposition is independent of time. In particular, $A(k_c)$ and $N(k_c)$ are independent of time. 
Second, the vortices at the same scale have a constant intensity, and the intensities at each scale are all represented by $A(k_c)$. 
Third, this decomposition does not depend on the kinematic viscosity $\nu$.

These assumptions are partially justified by our DNS.
We have evaluated the total length $L(k)$ of tubular vortices at each scale, corresponding to the left-hand side of \eqref{decomp}. 
The temporal stationarity of $L(k)$ in figure \ref{fig:T-dep} implies that the right-hand side of \eqref{decomp} is also stationary.
In particular, it means that $N(k_c)$ is independent of time if we consider the vortices at the same scale have a characteristic length proportional to $\ell={k_c}^{-1}$.
Note that it is impossible to discuss the time dependence of $A(k_c)$ with this calculation alone. 
However, the assumption on $A(k_c)$ is regarded as a way to simplify the problem, 
since only the structure of coherent vortices is taken into account in Ref.~\cite{Yoneda-2021}. 
As already shown in figure \ref{fig:L}, the robustness for $\nu$ ($R_\lambda$ in our DNS) is also verified.

In our DNS, it is shown that $L(k_c)$ obeys the power law (\ref{eq:2.0}) irrespective of time and the Reynolds number. 
These results are consistent with the derivation of the $-5/3$ law in Ref.~\cite{Yoneda-2021}. 
If we want to evaluate the deviation from the $-5/3$ law in more detail, we need to consider the strength of vortices, which we have not dealt with so far. 
This issue is discussed in the following subsections.

\subsection{Self-similar hierarchy of strong vortices}

The numerical results in \S 4 show that the number density of coherent tubular vortices at different scales is independent of the scale. This means that no intermittency effects are observed in the number density. However, it is well known that the energy dissipation rate in turbulence is intermittent. It has been considered that the spatial intermittency of the energy dissipation rate originates from the accumulation of the energy flux through a scale-by-scale energy cascade process. Here, we investigate this intermittency effect of the energy flux by evaluating the activity of the energy cascade in terms of the band-pass-filtered vorticity $\omega_\ell$ in the range $[k_c/2,k_c)$ with $\ell={k_c}^{-1}$ on each vortex axis. In practice, we add another criterion to the identification method described in \S 2; we discard the candidates at which the enstrophy $\Omega_\ell=|\omega_\ell|^2$ is smaller than  $\alpha\overline{\Omega_\ell}$, where $\overline{\Omega_\ell}$ is the spatial average of $\Omega_\ell$. Then, we estimate the total length $L^{(\alpha)}(k_c)$ of the axes of those strong vortices.

First, we show $L^{(\alpha)}(k_c)$ with $\alpha=1$, $2$, $3$ and $4$ at three different instances of the turbulence at $\langle R_\lambda\rangle=350$ in figure \ref{fig:w2}(a). Note that we plot $L^{(\alpha)}/{k_c}^{1.92}$. We have heuristically found this exponent 1.92 so that we can demonstrate $L^{(2)}(k_c)\propto{k_c}^{1.92}$. This figure shows that the hierarchy of stronger tubular vortices is indeed intermittent, though the dimension $1.92+1=2.92$ of the hierarchy is not far from $3$. It is also clear that $L^{(\alpha)}(k_c)$ is also independent of time as in $L(k_c)$ (figure \ref{fig:L}). Next, we plot $L^{(\alpha)}(k_c)$ at different Reynolds numbers in figure \ref{fig:w2}(b) to show that $L^{(\alpha)}(k_c)$ is independent of $R_\lambda$ except the low-wavenumber range. Incidentally, carefully looking at the curves for different $\alpha$ in figure \ref{fig:w2}, we notice that the scaling range expands to a higher wavenumber range for larger $\alpha$. This reminds us of multifractal models of intermittency, in which the upper bound of the inertial range depends on the order of the structure function.

The results in figure \ref{fig:w2} indeed indicate intermittency effects, but the effects are observed in the dissipation range, or more precisely, in a lower part of the B range ($0.02\langle\eta\rangle^{-1}\lesssim k_c\lesssim 0.1\langle\eta\rangle^{-1}$). In other words, the fractal nature observed in figure \ref{fig:w2} does not reflect spatial intermittency in the inertial range (i.e.~the T range with $k\lesssim0.02\langle\eta\rangle^{-1}$).

To further investigate intermittency, we estimate the probability density function (PDF) of the band-pass-filtered velocity components $\Delta u_\ell$ on vortex axes. If we normalize $\Delta u_\ell$ by $\ell^{1/3}$, then the PDF seems identical in the T range (figure is omitted), though the range is quite narrow in our DNS. In contrast, in the B range, the tail of the PDF is more pronounced for larger $k_c$ (i.e.~smaller $\ell$). Therefore, intermittency effects in the inertial range are rather weak in turbulence at the {\it moderate} Reynolds numbers ($R_\lambda=500$--$600$) and those in the dissipation range are much more significant.

\begin{figure}
\begin{center}
\includegraphics[bb=0 0 592 244,width=0.9\textwidth]{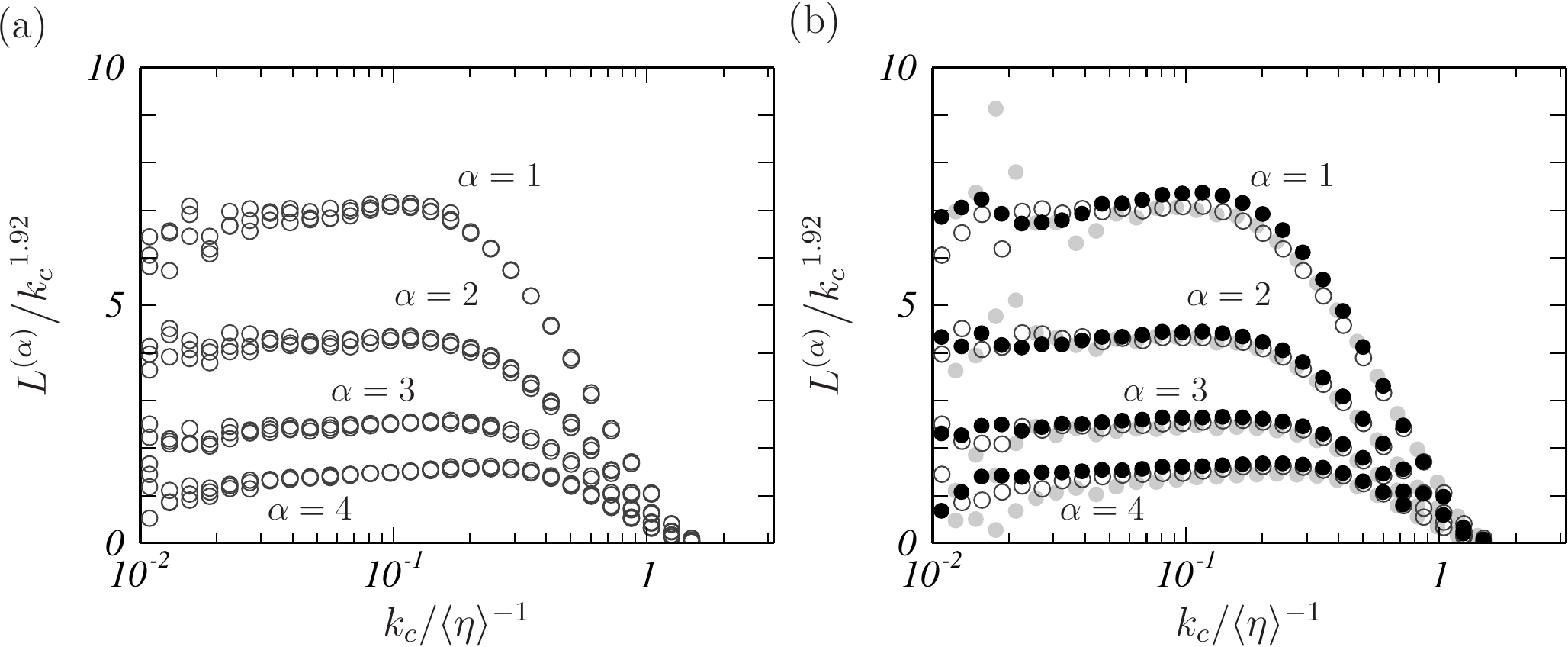}
\end{center}
\caption{(a) The total length $L^{(\alpha)}(k_c)$ ($\alpha=1$, $2$, $3$ and $4$) of strong vortices with enstrophy being $\alpha$ times larger than its spatial mean. To show the scaling $L^{(2)}(k_c)\propto {k_c}^{1.92}$, we plot $L^{(\alpha)}(k_c)/{k_c}^{1.92}$ as functions of $k_c$. The results at three different instances of turbulence driven by $f_I$ at the Reynolds number $\langle R_\lambda\rangle=350$. (b) The Reynolds-number dependence of $L^{(\alpha)}$. Different symbols correspond to different $\langle R_\lambda\rangle$; see figure \ref{fig:L}.}
\label{fig:w2}
\end{figure}

\subsection{Multifractal model}

We have developed arguments on the energy cascade in terms of coherent vortices in our previous studies \cite{Goto-2008,Goto-2012,Goto-2017,Motoori-2019,Motoori-2021,Yoneda-2021}, and we have objectively identified the hierarchy of their axes in the present study. Therefore, we may construct a concrete cascade model to explain the spatial intermittency of the energy flux and the dissipation rate. Here, we discuss the probabilistic formulation of the multifractal model (see \S 8.5.4 of Ref.~\cite{Frisch-1995}). First, suppose that we can estimate the velocity difference $\Delta u_\ell$ on each vortex axis in a given scale $\ell={k_c}^{-1}$. In practice, this can be done by an interpolation of the velocity field the Fourier-filtered velocity field in the wavenumber band (\ref{chi}). Then, we can estimate the $p$-th order moment of $\Delta u_\ell$ according to the multifractal analysis of the hierarchy of vortex axes as follows. In a given scale $\ell$ in the hierarchy, we denote by $dP_\ell(h)$ the probability density for $\Delta u_\ell$ to be $v_0(\ell/\ell_0)^h$. Here, $v_0$ and $\ell_0$ are constants and $h$ denotes the H\"older exponent. Then, we can estimate the $p$-th order moment of $\Delta u_\ell$ as
\begin{equation}
 \label{eq:du^p}
 \overline{\Delta u_\ell^p}=\int {v_0}^p\:\left(\frac{\ell}{\ell_0}\right)^{hp}\:dP_{\ell}(h)
 \:.
\end{equation}
It is important that we may numerically estimate $dP_\ell(h)$ by evaluating the total length $L_h(\ell)$ of the parts of vortex axes where the velocity difference $\Delta u_\ell$ is $v_0(\ell/\ell_0)^h$. Here, we expect that $L_h$ obeys a power law,
$
 L_h(\ell)\propto {k_c}^{\widetilde{D}(h)}\propto {\ell}^{-\widetilde{D}(h)}
$,
when $\ell$ is in the inertial range. Then, $dP_\ell(h)$ is expressed as
\begin{equation}
 \label{eq:dP}
 dP_\ell(h)
 \propto
 A(h)\:
 \left(\frac{\ell}{\ell_0}\right)^{3-D(h)}
 dh
 =
 A(h)\:
 \left(\frac{\ell}{\ell_0}\right)^{2-\widetilde{D}(h)}
 dh
\end{equation}
where $D(h)=\widetilde{D}(h)+1$ and $A(h)$ is a constant depending on $h$. Here, we have assumed again that the radius of vortex tubes are independent of their intensity and  proportional to $\ell={k_c}^{-1}$. Thus, substituting (\ref{eq:dP}) to (\ref{eq:du^p}) and using the steepest descent method, we can express the exponent $\zeta_p$ of $\overline{\Delta u_\ell^p}\propto\ell^{\zeta_p}$ in terms of $\widetilde{D}(h)$, which may be evaluated by using DNS data.

Since we have shown the power-law behavior (figures \ref{fig:L} and \ref{fig:w2}) of the total length $L$ and the length of strong parts of the vortex axes, we expect that we can evaluate $D(h)$ by estimating $L_h$ in a similar manner. This kind of analysis is attractive because we can reveal the physical origin of intermittency in terms of the realistic coherent structures. Unfortunately, however, this is impossible with our DNS data. As discussed in \S 3, in order to discuss the intermittency effects (e.g.~the deviation $\mu$ from the Kolmogorov spectrum), we have to evaluate $D(h)$ using the data in the T range (i.e.~$k\lesssim0.03\langle\eta\rangle^{-1}$). However, as discussed in the previous subsection, the fractal nature observed in figure \ref{fig:w2} is mainly in the B range, which is contaminated by viscous effects. In other words, the T range in our DNS data at $R_\lambda\approx500$--$600$ is too narrow to evaluate intermittency effects in the inertial range.

Detailed multifractal analysis of the T range with DNS data at higher Reynolds numbers is a future interesting study. It is also an interesting issue to understand the physical and mathematical origin of the boundary between F and T ranges of $E(k)$ around $k\approx0.005\langle\eta\rangle^{-1}$ \cite{Ishihara-2016}.

\section{Conclusion}

In our previous study \cite{Yoneda-2021}, by using the notion of the hierarchy of coherent tubular vortices, we mathematically reformulated the energy cascade process and derived the $-5/3$ power law of the energy spectrum from the Navier-Stokes equation without directly using the Kolmogorov similarity hypothesis. In the present article, we have numerically examined the assumptions in the formulation. To this end, we have applied the band-pass filter (\ref{chi}) to the turbulent velocity fields obtained by DNS, and applied the low-pressure method \cite{Miura-1997,Kida-1998,Goto-2017} to the scale-decomposed fields so that we can objectively identify the skeletal structures of coherent tubular vortices at each scale. The identified hierarchy (figure \ref{fig:axes}) of vortex axes is self-similar. To evaluate the dimension of the hierarchy, we have estimated the total length $L(k_c)$ of the vortex axes in the wavenumber band $[k_c/2,k_c)$. As shown in figure \ref{fig:L}, $L(k_c)$ obeys the power law (\ref{eq:2.0}), even without time averaging, in a significantly wide range of $k_c$. More precisely, this power law holds in the wavenumber range (\ref{eq:range}); namely, not only the T range (within the inertial range) but also a lower part of the B range (within the dissipation range), where the energy spectrum is accompanied by a bump. See figure \ref{fig:E} for the definitions of the T and B ranges. This result implies that there exists the energy cascade process due to vortex stretching even in the dissipation range, though the vorticity is attenuated by the viscosity in the range. In fact, this result is also consistent with the observation (figure 1(c) of Ref.~\cite{Yoneda-2021}) of the energy transfer in the B range. In other words, the number of child vortices in the cascading process is constant irrespective of the scale in the range (\ref{eq:range}). We have also shown that $L$ is robust and independent of Reynolds number and forcing (figure \ref{fig:L}), which implies the universality of turbulence. It is particularly important for the mathematical formulation of the energy cascade that $L$ is time independent (figure \ref{fig:T-dep}) even when the kinetic energy and its dissipation rate significantly fluctuate. This result supports the assumption in Ref.~\cite{Yoneda-2021}; see \S 4(a). The results in figure \ref{fig:L} also show that the intermittency of the energy dissipation rate is not due to the increase of volume fraction of smaller scale vortices but due to the accumulation of the energy flux in smaller scales; see \S 4(b). Therefore, we may further improve our mathematical formulation of the energy cascade \cite{Yoneda-2021}, by taking into account the multifractal nature (\S 4(c)), to describe the deviation of the energy spectrum from the $-5/3$ law.}

\appendix

\section{Practical procedure to obtain a candidate}

Here, $\bm{e}^{(i)}$ denotes the unit eigenvector associated with $\lambda^{(i)}$ and symbols in the main text are shown in vector notation. We assume $\lambda^{(2)} > 0$. For the grid point $\bm{a}$, we put 
$
\bm{c} = \bm{a} + \bm{\xi}
$.
Note that $\bm{\nabla} p_c(c)$ is parallel to $\bm{e}^{(3)}$. 
Since $p_c$ is approximated by its second-order Taylor polynomial, we have
\begin{eqnarray}
  \left\{
    \begin{array}{l}
       0 = \bm{e}^{(1)} \cdot \bm{\nabla}p_c(a) + \lambda^{(1)} \bm{e}^{(1)}\cdot \bm{\xi},
\\
       0 = \bm{e}^{(2)} \cdot \bm{\nabla}p_c(a) + \lambda^{(2)} \bm{e}^{(2)}\cdot \bm{\xi},
\\
       0 = \bm{e}^{(3)}\cdot \bm{\xi}.
    \end{array}
  \right.
\end{eqnarray}
Hence, in order to obtain $\bm{c}$, it suffices to solve the following equation:
\begin{equation}
 \left[
  \bm{e}^{(1)},
  \bm{e}^{(2)},
  \bm{e}^{(3)}
 \right]
 \left[
  \begin{array}{c}
   \xi_1 \\
   \xi_2 \\
   \xi_3 
  \end{array}
 \right]
 = 
 \left[
  -\frac{\bm{e}^{(1)}\cdot\bm{\nabla}p_c}{\lambda_1},
  -\frac{\bm{e}^{(2)}\cdot\bm{\nabla}p_c}{\lambda_2},
  0
 \right].
\end{equation}






\vspace{0.5cm}
\noindent
{\bf Acknowledgments.}\ 
Research of SG was partly supported by the JSPS Grants-in-Aid for Scientific Research  20H02068 and 20K20973.
TY  was partly supported by the JSPS Grants-in-Aid for Scientific
Research  20H01819.





\begin{thebibliography}{99}


\bibitem{Corrsin-1943}
Corrsin S.
1943. 
Investigations of flow in an axially symmetric heated jet of air. 
{\em NACA Adv. ConI Rep.}, 3123.

\bibitem{Hamilton-1995}
Hamilton JM, Kim J,  Waleffe F.
1995. 
Regeneration mechanisms of near-wall turbulence structures. 
{\em J. Fluid Mech.}, 287, 317-348.

\bibitem{Waleffe-1997}
Waleffe F. 
1997. 
On a self-sustaining process in shear flows. 
{\em Phys. Fluids}, 9, 883-900.

\bibitem{Tennekes-1972}
Tennekes H, Lumley JL
1972. 
A first course in turbulence. 
MIT press.


\bibitem{Frisch-1995} 
Frisch U. 
1995. 
Turbulence. 
Cambridge University Press, Cambridge. 

\bibitem{Kerr-1985}
Kerr RM. 
1985. 
Higher-order derivative correlations and the alignment of small-scale structures in isotropic numerical turbulence. 
{\em J. Fluid Mech.}, 153, 31.


\bibitem{Hussain-1986}
Hussain AKMF. 
1986. 
Coherent structures and turbulence. 
{\em J. Fluid Mech.}, 173, 303. 


\bibitem{Yamamoto-1988}
Yamamoto K, Hosokawa I.
1988. 
A decaying isotropic turbulence pursued by the spectral method. 
{\em J. Phys. Soc. Jpn}, 57, 1532-1535.


\bibitem{Melander-1993}
Melander MV, Hussain F. 
1993. 
Coupling between a coherent structure and fine-scale turbulence. 
{\em Phys. Rev. E}, 48, 2669-2689


\bibitem{Lundgren-1982}
Lundgren TS. 
1982. 
Strained spiral vortex model for turbulent fine structure. 
{\em Phys. Fluids}, 25, 2193-2203


\bibitem{Horiuti-2008}
Horiuti K, Fujisawa T. 
2008. 
The multi-mode stretched spiral vortex in homogeneous isotropic turbulence. 
{\em J. Fluid Mech.}, 595, 341-366.




\bibitem{Kerr-2013}
Kerr RM. 
2013. 
Swirling, turbulent vortex rings formed from a chain reaction of reconnection events. 
{\em Phys. Fluids}, 25, 065101.

\bibitem{Cardesa-2017}
Cardesa JI, Vela-Martin A, Jim\'enez J.
2017.
The turbulent cascade in five dimensions.
{\em Science}, 357, 782.

\bibitem{Doan-2018}
Doan NAK, Swaminathan N, Davidson PA, Tanahashi M.
2018.
Scale locality of the energy cascade using real space quantities.
{\em Phys. Rev. Fluids}, 3, 084601.

\bibitem{Goto-2008}
Goto S. 
2008. 
A physical mechanism of the energy cascade in homogeneous isotropic turbulence. 
{\em J. Fluid Mech.}, 605, 355-366.  


\bibitem{Goto-2017}
Goto S, Saito Y, Kawahara G. 
2017. 
Hierarchy of antiparallel vortex tubes in spatially periodic turbulence at high Reynolds numbers. 
{\em Phys. Rev. Fluids}, 2, 064603.


\bibitem{Motoori-2019}
Motoori Y, Goto S. 
2019. 
Generation mechanism of a hierarchy of vortices in a turbulent boundary layer.
{\em J. Fluid Mech.}, 865, 1085-1109.


\bibitem{Motoori-2021}
Motoori Y, Goto S. 
2021. 
Hierarchy of coherent structures and real-space energy transfer in turbulent channel flow. 
{\em J. Fluid Mech.}, 911, A27.


\bibitem{Goto-2012}
Goto S. 
2012.  
Coherent structures and energy cascade in homogeneous turbulence. 
{\em Progr. Theor. Phys. Suppl.}, 195, 139-156.


\bibitem{Yoneda-2021}
Yoneda T, Goto S, Tsuruhashi T. 
2021. 
Mathematical reformulation of the Kolmogorov-Richardson energy cascade in terms of vortex stretching. 
{\em arXiv}, 2105.12459.


\bibitem{Kolmogorov-1941}
Kolmogorov AN. 
1941.
The local structure of turbulence in incompressible viscous fluid for very large Reynolds numbers. 
{\em Dokl. Akad. Nauk SSSR}, 30, 301-305.


\bibitem{Miura-1997}
Miura H, Kida S. 
1997. 
Identification of tubular vortices in turbulence. 
{\em J. Phys. Soc. Jpn.}, 66, 1331-1334.


\bibitem{Kida-1998}
Kida S, Miura H. 
1998. 
Swirl condition in low-pressure vortex. 
{\em J. Phys. Soc. Jpn.}, 67, 2166-2169.


\bibitem{Dimotakis-2000}
Dimotakis PE.
2000. 
The mixing transition in turbulent flows. 
{\em J. Fluid Mech.}, 409, 69-98


\bibitem{Leung-2012}
Leung T, Swaminathan N, Davidson PA. 
2012. 
Geometry and interaction of structures in homogeneous isotropic turbulence. 
{\em J. Fluid Mech.}, 710, 453-481.


\bibitem{Lozano-Duran-2016}
Lozano-Dur\'an A. Holzner M, Jim\'enez J. 
2016.
Multiscale analysis of the topological invariants in the logarithmic region of turbulent channels at a friction Reynolds number of 932. 
{\em J. Fluid Mech.}, 803, 356-394.


\bibitem{Lee-2017}
Lee J, Sung HJ, Zaki TA.
2017.
Signature of large-scale motions on turbulent/non-turbulent interface in boundary layers. 
{\em J. Fluid Mech.}, 819, 165-187.


\bibitem{Chong-1990}
Chong MS, Perry AE, Cantwell BJ. 
1990. 
A general classification of three‐dimensional flow fields. 
{\em Physic. Fluids A}, 2, 765-777.


\bibitem{Jeong-1995}
Jeong J, Hussain F. 
1995. 
On the identification of a vortex. 
{\em J. Fluid Mech.}, 285, 69-94.


\bibitem{Ishihara-2016}
Ishihara T, Morishita K., Yokokawa M, Uno A, Kaneda Y. 
2016. 
Energy spectrum in high-resolution direct numerical simulations of turbulence. 
{\em Phys. Rev. Fluids}, 1, 082403. 


\bibitem{Ohkitani-1992}
Ohkitani K., Kida S. 
1992. 
Triad interactions in a forced turbulence. 
{\em Phys. Fluids A}, 4, 794-802.


\bibitem{Yasuda-2014}
Yasuda T, Goto S, Kawahara G. 
2014. 
Quasi-cyclic evolution of turbulence driven by a steady force in a periodic cube. 
{\em Fluid Dyn. Res.}, 46, 061413. 






%
%


























\end{thebibliography}
\end{document}